\documentclass[runningheads,fleqn]{cl2emult}

\usepackage{makeidx}  
\usepackage{graphicx} 
\usepackage{subeqnar} 
\usepackage{multicol} 
\usepackage{cropmark} 
\usepackage{phys}     
\makeindex            

\begin{document}
\title*{Critical finite-size scaling with constraints:\protect\newline
Fisher renormalization revisited}
\toctitle{Critical finite-size scaling with constraints:
\protect\newline Fisher renormalization revisited}
%
%
\titlerunning{Finite-size scaling with constraints}
%
\author{Michael Krech}
\authorrunning{M. Krech}
%
%
\institute{Institut f\"ur Theoretische Physik, RWTH Aachen,
52056 Aachen, Germany}

\maketitle              

\begin{abstract}
The influence of a thermodynamic constraint on the critical finite-size
scaling behavior of three-dimensional Ising and XY models is analyzed by
Monte-Carlo simulations. Within the Ising universality class constraints
lead to Fisher renormalized critical exponents, which modify the asymptotic
form of the scaling arguments of the universal finite-size scaling functions.
Within the XY universality class constraints lead to very slowly decaying
corrections inside the scaling arguments, which are governed by
the specific heat exponent $\alpha$. If the modification of the
scaling arguments is properly taken into account in the scaling analysis
of the data, finite-size scaling functions are obtained, which are {\em
independent} of the constraint as anticipated by analytic theory.
\end{abstract}

\section{Introduction}
The theoretical investigation of classical spin systems has played a key role
in the understanding of phase transitions, critical behavior, scaling, and
universality \cite{Amit78,Parisi88}. In particular, the classical Ising, the
XY, and the Heisenberg model are the most relevant spin models in three
dimensions. Each of these simple models represents a universality class which,
\index{universality class!Ising} \index{universality class!XY}
\index{universality class!Heisenberg} apart from the spatial dimensionality
and the range of the interactions, is characterized by the number of
components of the order parameter \index{order parameter},
e.g, the magnetization in the case of ferromagnetic models. Real systems,
however, suffer from various kinds of imperfections, e.g., lattice defects,
impurities, or vacancies. In an experiment, which is designed to probe
critical behavior as a function of temperature, the presence of, say,
impurities on the lattice constitutes a thermodynamic constraint, because
in a given sample the impurity concentration will remain constant during
the temperature scans. According to the concepts of thermodynamics the impurity
concentration $n_i$ can be written as the derivative of the grand canonical
potential with respect to the chemical potential $\mu_i$ of the impurities,
where other parameters like the temperature and the volume of the system are
kept fixed. Now the question arises how the critical singularities in the
grand canonical potential are affected when the thermodynamic ensemble is
changed from 'fixed $\mu_i$' to 'fixed $n_i$', where the location of the
critical temperature $T_c$ depends on the particular values of $\mu_i$ or
$n_i$, respectively. The answer to this question has been given a long time
ago by Michael Fisher \cite{Fisher68}. Provided, that the critical singularites
have their usual form in the 'fixed $\mu_i$' ensemble, then the constraint
$n_i = const.$ \index{constraint} amounts to a reparameterization of the
reduced temperature $t = (T-T_c(n_i))/T_c(n_i)$ of the {\em constrained}
system in terms of the reduced temperature $\tau = (T-T_c(\mu_i))/T_c(\mu_i)$
of the {\em unconstrained} system according to \cite{Fisher68}
\begin{equation}
\label{ttau}
t = a \tau + b \tau |\tau|^{-\alpha} + \dots ,
\end{equation}
where $a$ and $b$ are nonuniversal constants and the dots indicate higher
order contributions. Apart from a linear term (\ref{ttau}) contains a
singular contribution which is characterized by the critical exponent
$1 - \alpha$ of the entropy density \index{entropy density}. Which of the
two terms in (\ref{ttau}) is the leading one for $t, \tau \to 0$ depends
on the sign of $\alpha$. Within the Ising universality class in $d = 3$
\index{universality class!Ising} dimensions $\alpha \simeq 0.109$
\cite{LGZJ85} so that $|\tau| \sim |t|^{1/(1-\alpha)}$ to leading order
and therefore the critical exponents \index{critical exponents}
$\beta$ (order parameter), $\gamma$ (susceptibility), and $\nu$ (correlation
length) of the unconstrained system undergo 'Fisher renormalization' in the
constrained system according to \cite{Fisher68}
\begin{equation}
\label{fren}
\beta \to \beta' = \beta / (1 - \alpha), \quad
\gamma \to \gamma' = \gamma / (1 - \alpha), \quad
\nu \to \nu' = \nu / (1 - \alpha) .
\end{equation}
The specific heat exponent $\alpha$ requires a more careful analysis,
because the specific heat is the temperature derivative of the entropy
which in addition to the 'renormalization' displayed in (\ref{fren})
causes a sign change
\begin{equation}
\label{frena}
\alpha \to \alpha' = -\alpha / (1 - \alpha) .
\end{equation}
Note that analytic background contributions to the entropy of the
unconstrained system become {\em singular} in the constrained system
due to the singularity in the reparameterization given by (\ref{ttau}).

Within the XY universality class in $d = 3$ the exponent $\alpha$ is
negative \index{universality class!XY} \cite{LGZJ85}, where probably the
best current estimate $\alpha \simeq -0.013$ is obtained from an
experiment on $^4$He near the superfluid transition \cite{LSNCI96}. For
negative $\alpha$ the linear term on the r.h.s. of (\ref{ttau}) is the
dominating one for $\tau \to 0$. However, the XY universality class
$\alpha$ is so small, that in practice the singular term in (\ref{ttau})
can never be neglected. Instead, the singular contribution to (\ref{ttau})
gives rise to very slowly decaying correction terms \index{correction terms}
which must not be confused with Wegner corrections to scaling
\index{corrections to scaling}. These correction terms have to be considered
in any scaling analysis in order to obtain correct values for the critical
exponents.

If the system is finite, which is neccessarily the case for any Monte - Carlo
simulation, all critical singularities are rounded, i.e., all quantities
are analytic functions of the thermodynamic parameters \cite{Fisher71} so
that a thermal singularity as shown in (\ref{ttau}) does not occur.
Critical finite-size rounding effects in, e.g., a cubic box
\index{finite-size rounding} \index{finite-size scaling}
$L^d$ are captured by {\em universal} finite-size scaling functions
\index{finite-size scaling!function} \cite{Fisher71,Barber83} which restore
all critical singularities in the limit $L \to \infty$. Following the line
of argument in \cite{Fisher68}, (\ref{ttau}) then has to be replaced by
\begin{equation}
\label{ttaufL}
t = a \tau + \tau |\tau|^{-\alpha} f(\tau L^{1/\nu}) ,
\end{equation}
where $f(x)$ is the finite-size scaling function of the entropy density
and $x = \tau L^{1/\nu}$ is a convenient choice of its scaling argument.
For $\tau \to 0$ at finite $L$ the singular prefactor of $f(x)$ in
(\ref{ttaufL}) must be cancelled so that one has $f(x) = A |x|^{\alpha} +
\dots$ in the limit $x \to 0$, where $A$ is a nonuniversal constant such
that $f(x)/A$ is a {\em universal} function of its argument. To
leading order in $\tau$ the reparameterization of the reduced temperature
$t$ of the constrained system is therefore {\em linear} in the reduced
temperature $\tau$ of the unconstrained system and one finds
\begin{equation}
\label{ttauL}
t = \tau (a + A L^{\alpha/\nu}).
\end{equation}
According to (\ref{ttauL}) the finite-size scaling argument $x$ in the
constrained system is given by
\begin{equation}
\label{x}
x = \tau L^{1/\nu} = t L^{1/\nu} / (a + A L^{\alpha/\nu}),
\end{equation}
where the {\em shape} of the finite-size scaling functions is maintained
\cite{VD}, i.e., the presence of the constraint {\em only} affects the form
of the scaling argument $x$. For $\alpha > 0$ (\ref{x}) asymptotically
reduces to $x = t L^{1/\nu'}/A$ for large $L$ in accordance with Fisher
renormalization (see (\ref{fren})). For $\alpha < 0$ (\ref{x})
captures the aforementioned slowly decaying corrections to the asymptotic
critical behavior in the XY universality class when a thermodynamic constraint
is present. Note that $A > 0$ for the Ising universality class and that
$A < 0$ for the XY universality class.

In the remainder of this paper a simple spin model is introduced which can
be efficiently simulated with existing Monte - Carlo algorithms both with
and without constraints in three dimensions. For the Ising and the XY
version of the model finite-size scaling according to (\ref{x}) is
tested for the modulus of the order parameter, the susceptibility, and the
specific heat.

\section{Model and simulation method}
The model system which is investigated here can be described as an $O(N-1)$
symmetric classical 'planar' ferromagnet in a transverse magnetic field.
The model Hamiltonian reads \index{planar ferromagnet}
\begin{equation}
\label{H}
{\cal H} = -J \sum_{\langle i j \rangle}
\sum_{x=1}^{N-1} S_i^x S_j^x - h \sum_i S_i^N,
\end{equation}
where $\langle i j \rangle$ denotes a nearest neighbor pair of spins on
a simple cubic lattice in $d = 3$ dimensions. The lattice contains $L$
lattice sites in each direction and in order to avoid surface effects
periodic boundary conditions are applied. Each spin ${\bf S}_i$ is a
classical spin with $N$ components $\vec{S}_i = \left(S_i^1,S_i^2,
\dots,S_i^N\right)$ with the normalization $|\vec{S}_i| = 1$ for each
lattice site $i$. The magnetic field $h$ in (\ref{H}) only acts on
the $N$-components of the spins which are not coupled by the exchange
interaction $J$. From the symmetry of the Hamiltonian it is obvious, that
the model belongs to the $O(N-1)$ universality class in $d = 3$, where
nonuniversal quantities like the critical temperature $T_c = T_c(h)$
depend on the strength of the transverse field $h$. Note that $T_c(h)$
is symmetric around $h = 0$ and decreases with increasing $h$, because
the spins become more and more aligned with the $N$-direction as $\pm h$ is
increased and due to the normalization condition the typical interaction
energy between pairs of spins is decreased.

The Hamiltonian given by (\ref{H}) defines the unconstrained model.
The constraint is imposed on the transverse magnetization $M$ in the form
\index{constraint} \index{transverse magnetization}
\begin{equation}
\label{Mconst}
M \equiv \sum_i S_i^N = const.,
\end{equation}
where the Hamiltonian of the constrained model is given by 
\begin{equation} 
\label{Hconst} 
{\cal H}_M = -J \sum_{\langle i j \rangle} 
\sum_{x=1}^{N-1} S_i^x S_j^x = {\cal H} + hM .
\end{equation} 
The critical temperature of the constrained model is a symmetric and
monotonically decreasing function of the prescribed transverse
magnetization $M$. The transverse field $h$ here plays the part of the
chemical potential $\mu_i$ of impurities (see Sect.~1) and the transverse
magnetization $M$ accordingly plays the part of the impurity concentration
$n_i$. It is also possible to implement $O(N)$ spin models with impurities
or vacancies with diffusion in order to mimic the situation discussed in
\cite{Fisher68}. However, the fact that the Hamiltonians given by
(\ref{H}) and (\ref{Hconst}) only require a single 'species' with a
single coupling constant leads to some simplifications in the algorithms.
Note that the symmetric constrained model $(M = 0)$ becomes equivalent to
the symmetric unconstrained model $(h = 0)$ for sufficiently large lattices.
In particular, both versions of the symmetric model have the same $T_c$.

\index{Monte-Carlo algorithm} \index{Monte-Carlo algorithm!hybrid}
The Monte-Carlo algorithm is chosen as a hybrid scheme, where each hybrid
Monte-Carlo step consists of 10 updates each of which can be one of the
following: one Metropolis sweep of the whole lattice, one single cluster
Wolff update \cite{Wolff89}, or one overrelaxation update of the whole
lattice \cite{CL94}, where the latter can only be applied for $N \geq 3$.
\index{Monte-Carlo algorithm!Metropolis}
The Metropolis algorithm updates the lattice sequentially and works in
the standard way for the unconstrained model. For the constrained model
the constraint $M = const.$ is observed locally by applying a Kawasaki
\index{Monte-Carlo algorithm!Kawasaki}
update dynamics for the $N^{th}$ components of the spins. For each lattice site
$i$ a nearest neighbor site $j$ is chosen randomly and a random amount of the
$N^{th}$ spin component is proposed for exchange such that $S_i^N + S_j^N$
remains constant. Then new spin components $(S_i^1,S_i^2,\dots ,S_i^{N-1})$
are proposed and the spin components $(S_j^1,S_j^2,\dots ,S_j^{N-1})$
are adjusted according to the spin normalization condition $|\vec{S}_i| =
|\vec{S}_j| = 1$. The local change $\Delta E$ of the configurational energy
is calculated according to (\ref{Hconst}). According to detailed balance
the proposed update is accepted with probability $p(\beta \Delta E)$, where
$\beta = 1/(k_B T)$. For our simulation we have chosen $p(x) = 1 / (\exp(x)
+ 1)$. Note that all updates must be proposed such that the new spin at
lattice site $i$ is taken from the uniform distribution on the unit sphere
in $N$ dimensions.

\index{Monte-Carlo algorithm!Wolff}
The Wolff algorithm also works the standard way \cite{Wolff89}, except that
{\em only} the first $N-1$ components of the spins are used for the cluster
growth, i.e., (\ref{H}) and (\ref{Hconst}) are treated as planar
ferromagnets. This means that a cluster update never changes the
$N^{th}$ component of any spin so that the Wolff algorithm is nonergodic in
this case. The cluster update is still a valid Monte-Carlo step in the
sense that it fulfills detailed balance, however, in order to provide a
valid Monte-Carlo algorithm it has to be used together with the Metropolis
\index{Monte-Carlo algorithm!hybrid}
algorithm described above in a hybrid fashion. The use of Wolff
updates allows us to take advantage of improved estimators \cite{Has90}
for magnetic quantities.

\index{Monte-Carlo algorithm!overrelaxation}
The overrelaxation part of the algorithm performs a microcanonical update
of the configuration in the following way. The local configurational energy
has the functional form of a scalar product of the spins, where according
to (\ref{H}) and (\ref{Hconst}) only the first $N-1$ components are
involved. With respect to the sum of its nearest neighbor spins each spin
has a transverse component in the $(S_i^1,S_i^2,\dots ,S_i^{N-1})$ plane
which does not enter the scalar product. The overrelaxation algorithm
scans the lattice sequentially, determines this transverse component for each
lattice site and flips its sign. This overrelaxation algorithm is similar
to the one used in \cite{CL94} and it quite efficiently decorrelates
subsequent configurations over a wider range of temperatures around the
critical point than the Wolff algorithm. However, overrelaxation can only be
applied for $N \geq 3$. In the following only the cases $N = 2$ (transverse
Ising) and $N = 3$ (transverse XY) are considered.

\index{Monte-Carlo algorithm!hybrid}
In a typical hybrid Monte-Carlo step we use three Metropolis {\em (M)},
seven single cluster Wolff {\em (C)} updates for $N = 2$ and three
Metropolis, five single cluster Wolff, and two overrelaxation updates
{\em (O)} for $N = 3$ in the critical region of the models. The inividual
updates are mixed automatically in the program so that the update sequences
{\em (M C C M C C M C C C)} for $N = 2$ and {\em (M C C M O C M C C O)} for
$N = 3$ are generated as one hybrid Monte-Carlo step. The shift register
generator R1279 given by the recursion relation $X_n = X_{n-p} \oplus
X_{n-q}$ for $(p,q) = (1279,1063)$ is used as the random number generator.
Generators like this are known to cause systematic errors in combination
with the Wolff algorithm \cite{cluerr}. However, for lags $(p,q)$ as large
as the ones used here these errors will be far smaller than typical
statistical errors. They are further reduced by the hybrid nature of our
algorithm due to the presence of several Metropolis updates in one hybrid
Monte-Carlo step \cite{AMFDPL}.

The hybrid Monte-Carlo scheme described above is employed for lattice sizes
$L$ between $L = 20$ and $L = 80$. For each system size and temperature
we perform at least 10 blocks of $10^3$ hybrid steps for equilibration
followed by $10^4$ hybrid steps for measurements. Each measurement block
yields an estimate for all static quantities of interest and from these we
obtain our final estimates and estimates of their statistical error following
standard procedures. At the critical point (see below) two or three times as
many updates have been performed. The integrated autocorrelation time of
\index{autocorrelation time}
the hybrid algorithm is determined by the autocorrelation function of the
energy or, equivalently, the modulus of the order parameter, which yield
the slowest modes for the Wolff algorithm. The autocorrelation times are
generally rather short, at the critical point they range from about 5 hybrid
Monte-Carlo steps for $L = 20$ to about 10 hybrid Monte-Carlo steps for
$L = 80$. The values for the equilibration and measurement periods
given above thus translate to roughly 100 and 1000 autocorrelation times,
respectively. In order to obtain the best statistics for magnetic quantities
a measurement is made after every hybrid Monte-Carlo step. All error bars
quoted in the following correspond to one standard deviation. The simulations
have been performed on the DEC alpha AXP workstation cluster at the Physics
Department and on HP RISC8000 workstations at the Computer Center of the RWTH
Aachen.

\section{Ising universality class}
For $N = 2$ (\ref{H}) and (\ref{Hconst}) describe a classical Ising
model in a (fixed) transverse field or with fixed transverse magnetization,
respectively. In the following we will only consider the constrained model
with the symmetric constraint $M = 0$ and with the constraint $m \equiv M/L^3
= 1/\sqrt{2}$. The symmetrically constrained model does not show Fisher
renormalization \cite{Fisher68} and we therefore use this case for tests of
the algorithm and for the production of data representative of the Ising
universality class in $d = 3$. The constraint $m = const. \neq 0$ breaks
the $S_i^N \to -S_i^N$ symmetry of the model and Fisher renormatization
should become visible within a certain temperature window around $T_c =
T_c(m)$. The width of this window is of course a nonuniversal property of
the model and in particular one expects this window to widen as $m$ is
increased. Due to the spin normalization condition $m$ cannot exceed unity
and one therefore also expects, that critical behavior becomes very difficult
to resolve numerically if $m$ is too close to its maximum value. Therefore,
$m = 1/\sqrt{2}$ is chosen as a compromise between good resolution in the
critical regime and a prominent Fisher renormalization effect.

\begin{figure}
\begin{center}
\includegraphics[width=0.8\textwidth]{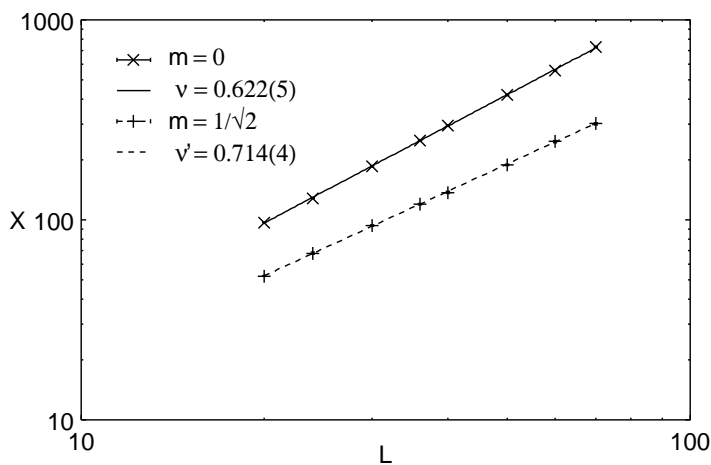}
\end{center}
\caption{Cumulant $X$ at the critical point for $m = 0$ ($\times$)
and $m = 1/\sqrt{2}$ (+). The solid and dashed lines display
power law fits to the data for $30 \leq L \leq 70$ for $m = 0$ and $m =
1/\sqrt{2}$, respectively}
\label{fig1}
\end{figure}

The critical temperatures $T_c(m=0)$ and $T_c(m=1/\sqrt{2})$ are determined
from temperature scans of the Binder cumulant ratio according to standard
procedures \cite{CFL93}. We obtain the following reduced critical coupling
constants $K_c(m) \equiv J / k_B T_c(m)$: \index{coupling constant!reduced}
\index{coupling constant!critical}
\begin{equation}
\label{KcI}
K_c(0) = 0.41638 \pm 0.00005 \quad \mbox{and} \quad
K_c(1/\sqrt{2}) = 0.6371 \pm 0.0001 .
\end{equation}
The corresponding estimates for the Binder cumulant ratio obtained for
$m = 0$ and $m = 1/\sqrt{2}$ agree with previous estimates obtained for
the Ising universality class within two standard deviations \cite{BLH95},
where for the latter choice of $m$ Wegner corrections to scaling
\index{corrections to scaling} are considerable and must be subtracted in
order to obtain a reliable estimate. In order to obtain an estimate for
the exponent $\nu$ which enters the finite-size scaling argument according
to (\ref{x}) the cumulant
\begin{equation}
\label{X}
X \equiv {\partial \over \partial T} \ln \langle \phi^2 \rangle
= {1 \over k_B T^2} \left(
{\langle \phi^2 {\cal H}_M \rangle \over \langle \phi^2 \rangle}
- \langle {\cal H}_M \rangle \right)
\end{equation}
has been measured, where $\phi = L^{-3}\sum_i S_i^1$ is the order parameter.
At the critical temperature $T_c(m)$ the scaling behavior $X \sim x/t$
is expected (see (\ref{x})). Corresponding numerical results for $m = 0$
and $m = 1/\sqrt{2}$ are displayed in Fig.\ref{fig1} on a double logarithmic
scale.
The data are compatible with simple power laws, where the exponents
$\nu = 0.622 \pm 0.005$ $(m = 0)$ and $\nu' = 0.714 \pm 0.004$ $(m =
1/\sqrt{2})$ have been obtained. Compared to the best currently known
estimate $\nu \simeq 0.630$ \cite{LGZJ85} the above estimate is too small
and only agrees with the theoretical value within two standard deviations.
A more thorough analysis shows that the discrepancy can be explained by
a mismatch of the order $5 \times 10^{-5}$ between the actual critical
temperature and the estimate used here (see (\ref{KcI})), which on the
other hand is of the same magnitude as the statistical error of $K_c(0)$.
The agreement between the above estimate for $\nu' = \nu / (1-\alpha)$ and
the theoretical value $\nu' \simeq 0.708$ \cite{LGZJ85} is better, however,
it may again be affected by a mismatch between the actual value
of $T_c(1/\sqrt{2})$ and the estimate used here. If the literature values
for $\nu$ and $\alpha$ are substituted in (\ref{x}), where $a$ and $A$
are used as fit parameters, $a/A \simeq 0.1$ is obtained which is small
enough to be ignored in the scaling analysis (see below).

The finite-size scaling analysis has been performed for several thermodynamic
quantities, in particular, the average modulus of the order parameter
$\langle |\phi| \rangle$, the susceptibilities \index{order parameter}
\index{susceptibility}
\begin{equation}
\label{Chi}
\chi_+ \equiv {L^3 \over k_B T} \langle \phi^2 \rangle , \quad
\chi_- \equiv {L^3 \over k_B T} \left( \langle \phi^2 \rangle
- \langle |\phi| \rangle^2 \right),
\end{equation}
\index{specific heat}
and the specific heat $C$. Data will only be shown for $\langle |\phi|\rangle$,
$\chi_-$, and $C$, because the finite-size scaling functions for $\langle
|\phi| \rangle$ and $\chi_+$ are very similar. According to finite-size
scaling theory it must be possible to callapse the data for all $m$ onto
one and the same curve, where two nonuniversal scaling factors are required
for each quantity. One scaling factor adjusts the magnitude of the scaling
argument $x$ (see (\ref{x})), the other adusts the absolute normalization
of the quantity. Note that the former saling factor must be the same for
{\em all} quantities. For $m = 0$ the scaling argument $x = t L^{1/\nu}$
is used, whereas for $m = 1/\sqrt{2}$ the choice $x = t L^{(1-\alpha)/\nu}
/ A$ has been made, where $A \simeq 1.1$ and the coefficient $a$ in
(\ref{x}) has been neglected. The exponents $\nu$ and $\alpha$ are taken
from the literature \cite{LGZJ85}.
\begin{figure}
\begin{center}
\includegraphics[width=0.8\textwidth]{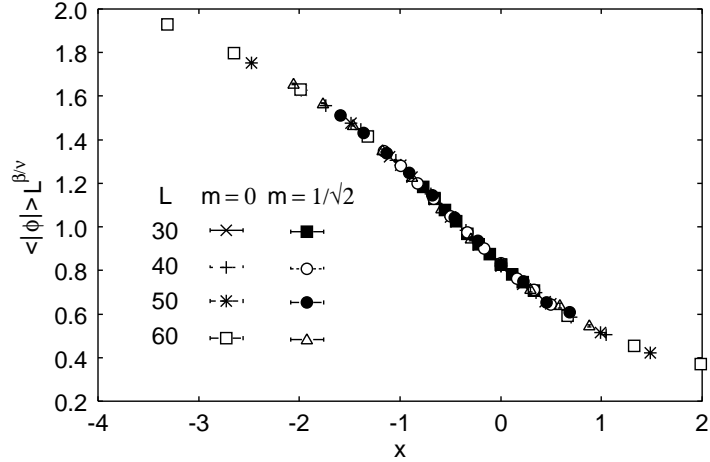}
\end{center}
\caption{Scaling plot of $\langle |\phi| \rangle$ for
$L = 30$, 40, 50, and 60 for $m = 0$ and $m = 1/\sqrt{2}$. The reduced
temperature $t$ has  been varied between $-0.007$ and 0.003. Statistical
errors are much smaller than the symbol sizes} 
\label{fig2}
\end{figure}
The scaling plot of $\langle |\phi|
\rangle$ is shown in Fig.\ref{fig2}, where the abolute normalization of the
data for $m = 1/\sqrt{2}$ can be adjusted to the $m = 0$ data by a scale
factor of $\sim 0.7$ as one would expect from simple mean field arguments.
As shown in Fig.\ref{fig2}, the espected data collapse can be reproduced
rather well, where the literature value for the exponent $\beta / \nu =
0.5168$ \cite{LGZJ85} has been used. \index{susceptibility}
The same holds for the susceptibility $\chi_-$ which is displayed in
Fig.\ref{fig3}, where $\gamma / \nu = 2 - \eta = 1.967$ is also taken form
\cite{LGZJ85}. The absolute magnitudes of $\chi_-/L^{\gamma/\nu}$
for $m = 0$ and $m = 1/\sqrt{2}$ are different by a factor of about $0.5$
which is in accordance with simple mean field arguments. Note that the
scaling function of $\chi_-$ has a maximum for $x < 0$ \cite{Dohm9596}.
\begin{figure}
\begin{center}
\includegraphics[width=0.8\textwidth]{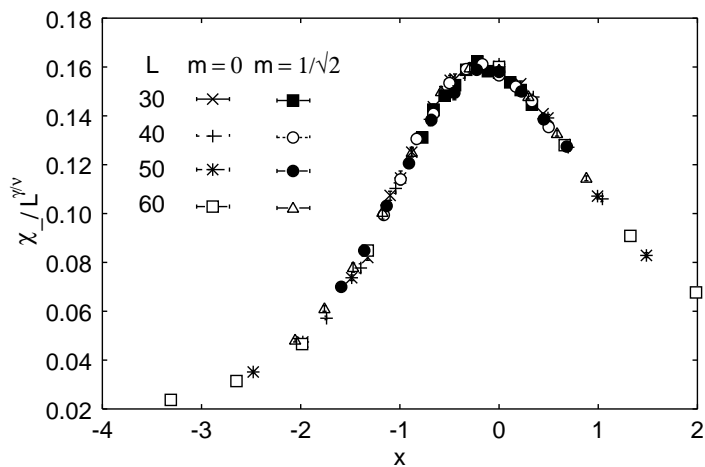}
\end{center}
\caption{Scaling plot of $\chi_-$ for $L = 30$, 40, 50, 
and 60 for $m = 0$ and $m = 1/\sqrt{2}$. The reduced temperature $t$ has 
been varied between $-0.007$ and 0.003. Statistical errors are much smaller
than the symbol sizes}
\label{fig3}
\end{figure}

\begin{figure}
\begin{center}
\includegraphics[width=0.8\textwidth]{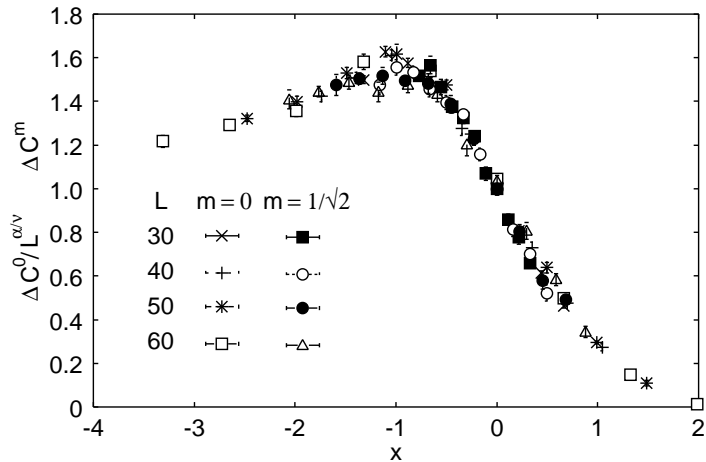}
\end{center}
\caption{Relative specific heats $\Delta C^0 /
L^{\alpha/\nu}$ for $m = 0$ and $\Delta C^m$ for $m = 1/\sqrt{2}$ for $L =
30$, 40, 50, and 60. The reduced temperature $t$ has been varied between
$-0.007$ and 0.003}
\label{fig4}
\end{figure}

The specific heat \index{specific heat} $C$ requires a somewhat different
treatment due to the fact that unlike the other quantities presented so
far the specific heat requires an {\em additive} renormalization within
renormalized field theory \cite{Dohm9596}. For the data analysis this means
that scaling can only be obtained after a suitable subtraction is applied
to the specific heat. One option to obtain scaling is to subtract the bulk
specific heat $C^0_b(t)$ at a reference reduced temperature $t = t_0$ which
are given by
\begin{equation}
\label{Cbt0}
C^0_b(t) = {A_\pm \over \alpha} |t|^{-\alpha} + B
\quad \mbox{and} \quad t_0 = (L/\xi^0_\pm)^{-1/\nu},
\end{equation}
where $A_\pm$ and $B$ are nonuniversal constants and $\xi^0_\pm$ is the
amplitude of the correlation length. The index $\pm$ refers to temperatures
above or below $T_c(m)$, respectively. The reference reduced temperature
chosen in (\ref{Cbt0}) is positive and therefore only $A_+$ and $\xi^0_+$
are needed. Specifically, the choice $\nu = 0.630$, $\alpha = 0.109$
\cite{LGZJ85}, $A_+ = 0.1552$, $B = -1.697$, and $\xi^0_+ = 0.495$
\cite{Dohm9596} guarantee scaling of the {\em relative} specific heat
$\Delta C^0 \equiv C - C^0_b(t_0)$ for the Ising model in $d = 3$. Note that
$\xi^0_+$ is measured in units of the lattice constant. For the data to be
analyzed here the subtraction defined by (\ref{Cbt0}) is only valid for
the case $m = 0$, where $\Delta C^0$ scales as $L^{\alpha/\nu}$. For
$m = 1/\sqrt{2}$ Fisher renormalization according to (\ref{ttau}) must be
\index{Fisher renormalization} applied to $C^0_b(t)$ in order to obtain the
correct form $C^m_b(t)$ of the subtraction. The result is
\begin{equation}
\label{Cbt0m}
C^m_b(t) = {A'_+ \over \alpha} |t|^{\alpha/(1-\alpha)} + B'
\quad \mbox{and} \quad t_0 = (L/\xi^0_\pm)^{(\alpha-1)/\nu},
\end{equation}
where $A'_+ = -0.1728$, $B' = 1.598$, and $\xi^0_+ = 0.495$ is not changed.
The scaling factor $L^{\alpha/\nu}$, which usually governs the finite-size
scaling of the specific heat, is cancelled here, i.e., one expects data
collapse for $\Delta C^0 / L^{\alpha/\nu}$ and $\Delta C^m \equiv C -
C^m_b(t_0)$ up to an overall scale factor of about $0.5$. The result of the
data analysis is shown in Fig.\ref{fig4}.
The data collapse reasonably
well onto a single curve except near the maximum of the scaling function,
where also the scatter of the individual data is substantial due to a few
bad samples for $L = 50$ and 60. However, there are also systematic
deviations from scaling in the data, because the maximum in the scaling
functions for the $m = 0$ data is more pronounced than in the $m = 1/\sqrt{2}$
data. These deviations could be due to enhanced Wegner corrections to scaling
for $m = 1/\sqrt{2}$ as compared to $m = 0$. \index{corrections to scaling}

\section{XY universality class}
\begin{figure}
\begin{center}
\includegraphics[width=0.8\textwidth]{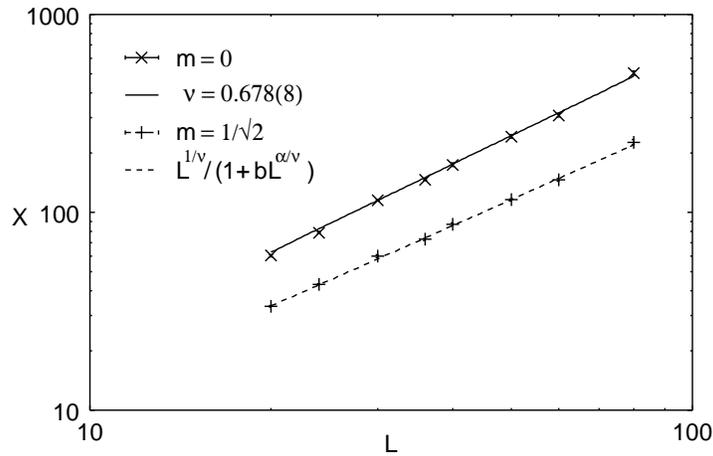} 
\end{center} 
\caption{Cumulant $X$ at the critical point for 
$m = 0$ (x) and $m = 1/\sqrt{2}$ (+). The solid and dashed lines display 
fits to the data for $30 \leq L \leq 80$ for $m = 0$ and $m = 
1/\sqrt{2}$, respectively (see main text)} 
\label{fig5} 
\end{figure}

For $N = 3$ (\ref{H}) and (\ref{Hconst}) describe a classical XY 
model in a (fixed) transverse field or with fixed transverse magnetization, 
respectively. As for the case $N = 2$ we will only consider the constrained
model with the symmetric constraint $m = 0$ and with the constraint $m =
1/\sqrt{2}$ in the following. The symmetrically constrained model is again
used for algorithmic tests and data production for the XY universality
class. The nonsymmetric constrained XY model does not show Fisher
renormalization, however, according to (\ref{x}) very slowly decaying
corrections to the asymptotic critical behavior are expected, which will be
discussed in the following. First, the cumulant $X$ defined by (\ref{X})
is evaluated at the critical point, which is given by the reduced coupling
constants \index{coupling constant!critical}
\begin{equation}
\label{KcXY}
K_c(0) = 0.6444 \pm 0.0001 \quad \mbox{and} \quad
K_c(1/\sqrt{2}) = 1.1126 \pm 0.0003 ,
\end{equation}
respectively. The result for $X$ is displayed in Fig.\ref{fig5}.
For $m = 0$ the data can be fitted by a power law $\sim L^{1/\nu}$, where
$\nu = 0.678 \pm 0.008$ is obtained which agrees with the best current
estimate $\nu = 0.671$ \cite{LGZJ85}. For $m = 1/\sqrt{2}$ the data can
also be fitted by a power law, however, the resulting exponent $\nu$ only
has the meaning of an effective exponent which does not fit into the XY
universality class. As shown in Fig.\ref{fig5} the expression $x/t$ according
to (\ref{x}) also yields a very good representation of the data where
the parameter $b$ in Fig.\ref{fig5} is given by $b = A/a = -0.941$. The
exponents used in the fit (XY universality class) are taken from
\cite{LGZJ85}. The value of the Binder cumulant found here agrees with
results reported in the literature for the standard (plane rotator) XY model
\cite{GH93}, however, Wegner corrections to scaling become quite substantial
for $m = 1/\sqrt{2}$.

\begin{figure}
\begin{center}
\includegraphics[width=0.8\textwidth]{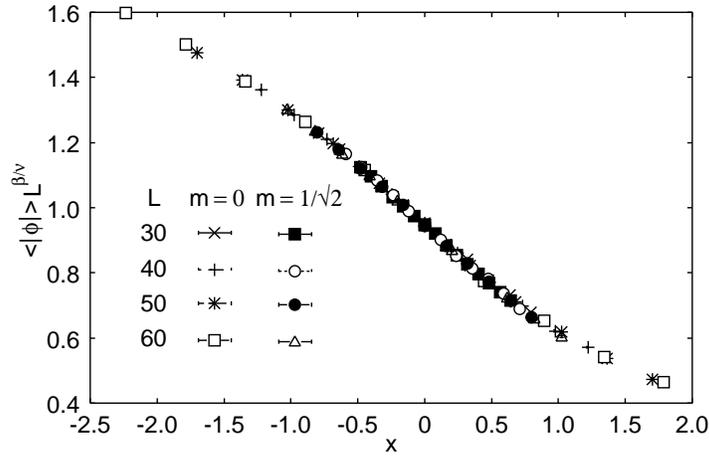}
\end{center}
\caption{Scaling plot of $\langle |\phi| \rangle$ for
$L = 30$, 40, 50, and 60 for $m = 0$ and $m = 1/\sqrt{2}$. The reduced
temperature $t$ has  been varied between $-0.005$ and 0.005. Statistical
errors are much smaller than the symbol sizes} 
\label{fig6}
\end{figure}

\begin{figure}
\begin{center}
\includegraphics[width=0.8\textwidth]{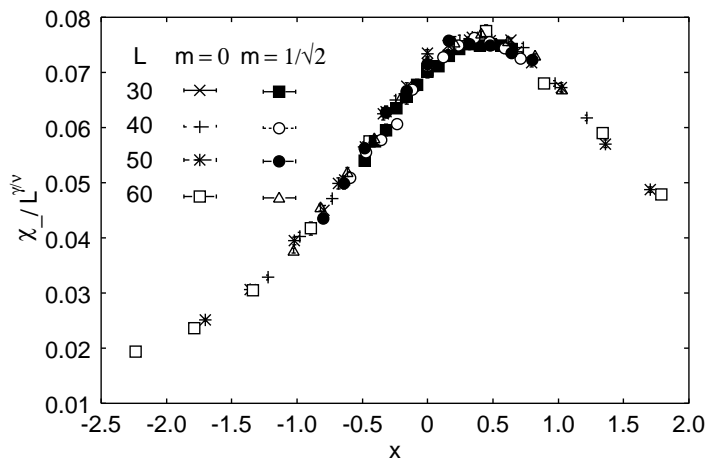}
\end{center}
\caption{Scaling plot of $\chi_-$ for $L = 30$, 40, 50, 
and 60 for $m = 0$ and $m = 1/\sqrt{2}$. The reduced temperature $t$ has 
been varied between $-0.005$ and 0.005. Statistical errors are much smaller
than the symbol sizes}
\label{fig7}
\end{figure}

In the following scaling analysis the finite-size scaling argument for the
case $m = 0$ takes its standard form $x = t L^{1/\nu}$ and for $m =
1/\sqrt{2}$ the combination $x = t L^{1/\nu} / (1+b L^{\alpha/\nu})$ for
$b = -0.941$ takes care of the slowly decaying correction terms to the
\index{correction terms} asymptotic critical behavior caused by
the very small and negative value of $\alpha$ in the XY universality class.
As in Sect.~3 we consider $\langle |\phi| \rangle$, $\chi_+$, $\chi_-$, and the
specific heat $C$ in the scaling analysis. The scaling functions for $\langle
|\phi| \rangle$ and $\chi_+$ again look very similar so that we do not
reproduce scaling plots for $\chi_+$ here. The result for $\langle |\phi|
\rangle$ is shown in Fig.\ref{fig6}, where the order parameter $\phi
\equiv L^{-3} \sum_i (S^1_i,S^2_i)$ has two components here.
\index{order parameter}
The data collapse very well onto a single curve. Note that the absolute
magnitudes of $\langle |\phi| \rangle$ for $m = 0$ and $m = 1/\sqrt{2}$
are again related by a factor of $\sim 0.7$ as suggested by mean-field
arguments. The values for the critical exponents $\nu = 0.671$ and $\beta
= 0.347$ are taken from the literature \cite{LGZJ85}. The susceptibility
\index{susceptibility} $\chi_-$ can be treated essentially as described
in Sect.3, where the exponent $\gamma / \nu = 2 - \eta = 1.965$ is taken
from \cite{LGZJ85}. The result of the scaling analysis is displayed in
Fig.\ref{fig7}.
The data do not collapse as well as in Fig.\ref{fig3}. Especially near the
maximum of the scaling function the scatter of the data is substantially
larger than in Fig.\ref{fig3}. Slight systematic deviations from scaling
for $m = 1/\sqrt{2}$ are observed which may again be due to enhanced Wegner
corrections to scaling as compared to $m = 0$. Note that contrary to
Fig.\ref{fig3} the scaling function of $\chi_-$ has a maximum for $x > 0$
in the XY universality class.

\index{specific heat}
The specific heat $C$ of the XY model also requires a subtraction before
scaling is obtained \cite{CDE95}. The subtraction $C^0_b(t_0)$ is again
used in the form given by (\ref{Cbt0}), where $A_+ = 0.42$, $\alpha =
-0.013$, $B = -A_+/\alpha$, and $\xi^0_+ = 1.0$ are used here which differ
somewhat from the choices made for the standard XY model in \cite{CDE95}.
It turns out, that the quality of the data collapse for the relative specific
heat $\Delta C^0 = C - C^0_b(t_0)$ for $m = 0$ is rather insensitive to the
choice of $\xi^0_+$. The form of the subtraction $C^m_b(t_0)$ for $m =
1/\sqrt{2}$ requires a little analysis in order to include the slowly varying
corrections to the asymptotic behavior coming from (\ref{ttau}). One
obtains the approximate form
\begin{equation}
\label{Cbt0XY}
C^m_b(t_0) = {A_+ / \alpha \over 1 + c t_0^{-\alpha}} \left[ \left({t_0 \over
1 + c t_0^{-\alpha}}\right)^{-\alpha} - 1\right], \quad t_0 = L^{-1/\nu}
(1 + b L^{\alpha/\nu}),
\end{equation}
where $b = -0.941$ (see Fig.\ref{fig5}), $A_+ = 0.42$ as before, and $c
\simeq 2.0$ for optimal data collapse.
\begin{figure}
\begin{center}
\includegraphics[width=0.8\textwidth]{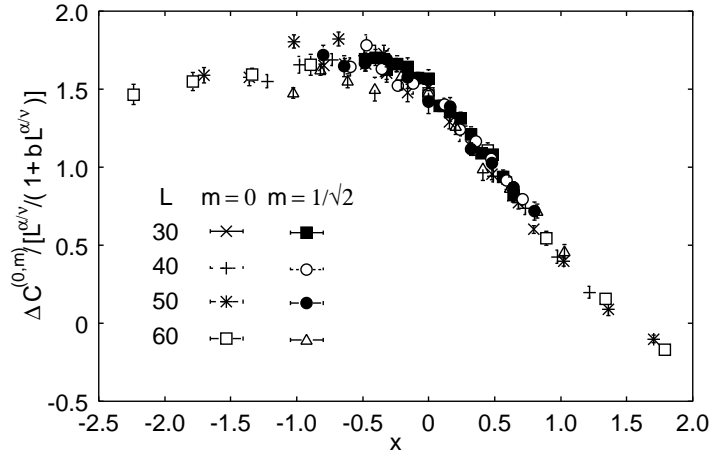}
\end{center}
\caption{Relative specific heats $\Delta C^{(0,m)} / \left[ L^{\alpha/\nu}
/ (1 + bL^{\alpha/\nu}) \right]$ with $b = 0$ for $m = 0$ and $b = -0.941$
for $m = 1/\sqrt{2}$ for $L = 30$, 40, 50, and 60. The reduced
temperature $t$ has been varied between $-0.005$ and 0.005}
\label{fig8}
\end{figure}
The resulting scaling plot of $\Delta C^0$ and $\Delta C^m \equiv C -
C^m_b(t_0)$ is displayed in Fig.\ref{fig8}.
The overall shape of the scaling function is similar to the one shown in
Fig.\ref{fig4}. However, the scatter of the data near the maximum is so
strong, that data collapse cannot be obtained in this region. In part this
deficiency in the data may be due to the presence of 'bad' samples, e.g,
for $L = 50$ and $m = 0$ at $t = -0.003$ and $t = -0.002$ and for $L = 60$ and
$m = 1/\sqrt{2}$ at $t = -0.005$ and $t = -0.003$. Apart from that deviations
from scaling as in Fig.\ref{fig4} may be present which are due to an
enhancement of Wegner corrections to scaling \index{corrections to scaling}
for $m = 1/\sqrt{2}$ as compared to the case $m = 0$. However,
Figs.\ref{fig6} - \ref{fig8} confirm, that the choice of the scaling
variable $x$ given by (\ref{x}) captures the slowly decaying correction term
\index{correction terms} inside the scaling argument in an appropiate way and
that furthermore the finite-size scaling behavior of constrained models in
the Ising and the XY universality class can be treated on the same footing.

\section{Summary and conclusions}
The influence of constraints on the critical finite-size scaling behavior
of Ising and XY models has been investigated by Monte-Carlo simulations
of $O(N-1)$ planar ferromagnets with fixed transverse magnetization. The
theoretical idea that only the form of the scaling argument is modified,
whereas the shape of the universal scaling functions remains unchanged
is verified within the statistical uncertainty of the data for the modulus
of the order parameter, the susceptibilites $\chi_+$ (not shown) and $\chi_-$,
and the specific heat. The form of the scaling argument used here allows
to deal with critical finite-size effects in constrained Ising and XY models
on the same footing, where constrained Heisenberg models can be included as
well. Within the Ising universality class the finite-size behavior is
consistent with the Fisher renormalization of critical exponents. In the XY
universality class slowly decaying corrections to the asymptotic critical
behavior are generated which are captured systematically by the analytic
form of the scaling argument. The treatment of these corrections within the
XY universality class may serve as a paradigm for the finite-size scaling
analysis of dynamic quantities, where the constraints imposed here reappear
as conserved quantities which are statically or dynamically coupled to the
order parameter. These corrections may also be important for the interpretation
of spin dynamics data for planar ferromagnets, where the energy of the system
is conserved during the simulated time evolution of spin models similar to the
ones investigated here.
\bigskip

\subsection*{Acknowledgments}
The author gratefully acknowledges many helpful discussions with V. Dohm
and financial support of this work through the Heisenberg program of the
Deutsche Forschungsgemeinschaft.

\clearpage
\addcontentsline{toc}{section}{Index}
\flushbottom
\printindex


\noindent

\begin{thebibliography}{99}
\addcontentsline{toc}{section}{References}
\bibitem{Amit78}
 Amit D. J. (1978) Field Theory, the Renormalization Group, and Critical
 Phenomena, McGraw-Hill, New York
\bibitem{Parisi88}
 Parisi G. (1988) Statistical Field Theory, Addison-Wesley, Wokingham
\bibitem{Fisher68}
 Fisher M. E. (1968) Phys. Rev. {\bf 176}, 257
\bibitem{LGZJ85}
 Le Guillou J. C., Zinn-Justin J. (1985) J. Phys. (France) Lett. {\bf 46},
 L137; Guida~R., Zinn-Justin~J. (1998) J. Phys. A {\bf 31}, 8103
\bibitem{LSNCI96}
 Lipa J. A., Swanson D. R., Nissen J. A., Chui T. C. P., Israelsson U. E.
 (1996) Phys, Rev. Lett. {\bf 76}, 944
\bibitem{Fisher71}
 Fisher M. E. (1971) in: Green M. S. (Ed.) Proceedings of the 1970 Enrico
 Fermi School of Physics, Varenna, Italy, Course No. LI Academic, New, York,
 p.1; Fisher~M.~E., Barber~M.~N. (1972) Phys. Rev. Lett. {\bf 28}, 1516
\bibitem{Barber83}
 Barber M. N. (1983) in: Domb C., Lebowitz J. L. (Eds.) Phase Transitions
 and Critical Phenomena, Vol. 8, Academic, New York, p. 145; Privman~V. (1990)
 in: Privman V. (Ed.) Finite Size Scaling and Numerical Simulation of
 Statistical Systems, World Scientific, Singapore
\bibitem{VD}
 Dohm V. (1998) private communication
\bibitem{Wolff89} 
 Wolff U. (1989) Phys. Rev. Lett. {\bf 62}, 361
\bibitem{CL94} 
 Chen K., Landau D. P. (1994) Phys. Rev. B {\bf 49}, 3266
\bibitem{Has90}
 Hasenbusch M. (1990) Nucl. Phys. {\bf B333}, 581
\bibitem{cluerr}
 Ferrenberg A. M., Landau D. P., Wong Y. J. (1992) Phys. Rev. Lett. {\bf 69},
 3382; Shchur~L.~N., Bl\"othe~H.~W.~J. (1997) Phys. Rev. E {\bf 55}, R4905
\bibitem{AMFDPL}
 Ferrenberg A. M., Landau D. P. unpublished
\bibitem{CFL93}
 Chen K., Ferrenberg A. M., Landau D. P. (1993) Phys. Rev. B {\bf 48}, 3249
\bibitem{BLH95}
 Bl\"ote H. W., Luijten E., Heringa J. R. (1995) J. Phys. A {\bf 28}, 6289
\bibitem{Dohm9596}
 Esser A., Dohm V., Chen X. S. (1995) Physica A {\bf 222}, 355;
 Chen~X.~S., Dohm~V., Talapov~A.~L. (1996) Physica A {\bf 232}, 375
\bibitem{GH93}
 Gottlob A. P., Hasenbusch M. (1993) Physica A {\bf 201}, 593
\bibitem{CDE95}
 Chen X. S., Dohm V., Esser, A. (1995) J. Phys. I France {\bf 5}, 205;
 Schultka~N., Manousakis~E. (1995) Phys. Rev. B {\bf 52}, 7528
\end{thebibliography}
\end{document}